\documentclass[reprint, showpacs, amsmath, amssymb, aps]{revtex4-1}
\usepackage{graphicx,color}
\usepackage{epsfig}
\usepackage{amsmath}
\usepackage{bbm}
\usepackage{amsfonts}
\usepackage{amssymb}
\usepackage{amsthm}
\usepackage{verbatim}
\usepackage{CJK}
\usepackage[colorlinks, linkcolor=blue, citecolor=blue, urlcolor=blue, breaklinks=true]{hyperref}

\newcommand{\ket}[1]{\left\vert{#1}\right\rangle}
\newcommand{\bra}[1]{\left\langle{#1}\right\vert}
\newcommand{\abs}[1]{\left\vert{#1}\right\vert}

\begin{document}
\title{Mechanical cooling in the single-photon quadratical optomechanics}

\author{Wen-ju Gu}
\email[]{guwenju@yangtzeu.edu.cn}
\author{Zhen Yi}
\author{Li-hui Sun}
\author{Da-hai Xu}
\affiliation{Institute of Quantum Optics and Information Photonics, School of Physics and Optoelectronic Engineering, Yangtze University, Jingzhou 434023, China}
\date{\today}

\begin{abstract}
In the paper we study the nonlinear mechanical cooling processes in the intrinsic quadratically optomechanical coupling system without linearizing the optomechanical interaction. We apply the scattering theory to calculate the transition rates between different mechanical Fock states with the use of the resolvent of Hamiltonian, which allows for a direct identification of the underlying physical processes, where only even-phonon transitions are permitted and odd-phonon transitions are forbidden. We verify the feasibility of the approach by comparing the steady-state mean phonon number obtained from transition rates with the simulation of the full quantum master equation, and also discuss the analytical results in the weak coupling limit that coincide with two-phonon mechanical cooling processes. Furthermore, to evaluate the statistical properties of steady mechanical state, we respectively apply the Mandel Q parameter to show that the oscillator can be in a nonclassical mechanical states, and the phonon number fluctuations F to display that the even-phonon transitions favor to suppress the phonon number fluctuations compared to the linear coupling optomechanical system.

\end{abstract}


\maketitle
\section{Instruction}
Realization of cooling of mechanical motion within the vicinity of quantum regime is a key ingredient for the broad range of its applications, including quantum information processing (QIP)~\cite{Sh.barzanjeh@prl,V.Fiore@pra,C.Dong@sci}, high-precision measurement~\cite{J.Li@pr,T.J.Kippenberg@sci}, and probe of quantum behaviour of macroscopic system~\cite{A.A.Gangat@pra}, etc. Recently, theoretical and experimental studies are mainly focused on the linearized treatment of linear coupling optomechanical system, where the coupling term is proportional to the mechanical displacement $x$~\cite{I.Wilson-Rae@prl,J.D.Teufel@nat,W.Gu@pra}. However, there are some literatures have begun to generalize the investigations on the nonlinear mechanical cooling processes, such as second-sideband laser cooling of trapped ions~\cite{R.L.de Matos Filho} and cooling in the single-photon strong optomechanical coupling regime~\cite{A.Nunnenkamp@praR}, which can lead to nonthermal steady states, and even the nonclassical, sub-Poissonian states, and quantum systems far from thermal equilibrium hold great promise for the investigation of fundamental physics and the implementation of practical devices~\cite{M.Koch@prl,X.Lu@prl}.

By now a new type of optomechanical system, i.e., quadratic optomechanics, has been proposed, where the coupling term is proportional to square of mechanical displacement $x^2$, which is realizable in the setups of membrane-in-the-middle optomechanical system, ultracold atomic ensembles and superconducting electrical circuit systems~\cite{J.D.Thompson@nat,T.P.Purdy@prl,E.J.Kim@arxiv}. Distinct from the displacement of mechanical equilibrium position in the linear coupling optomechanics, the quadratic coupling will modify the resonant frequency of the mechanical oscillator~\cite{J.Q.Liao@pra,J.Q.Liao@SR,H.Shi@pra}. Thus the quadratic optomechanical system can be used for quantum non-demolition (QND) measurement of individual quantum jumps~\cite{A.A.Clerk@prl}. In addition, the robust stationary mechanical squeezing and electromagnetically induced transparency (EIT) from two-phonon processes in quadratic optomechanical system are also achievable~\cite{M.Asjad@pra,S.Huang@pra}. For the practical applications one should minimize the influence of thermal noise, and linearized approach on the cooling of quadratically coupled mechanical oscillator has been studied in Ref.~\cite{Z.J.Deng@pra}. With the advancement of experimental technology, now it will become attractive to theoretically investigate the nonlinear cooling processes induced by the intrinsic quadratical coupling. For example, utilizing avoided crossings effects in a multi-mode optical cavity containing a flexible dielectric membrane enables the significantly enhanced quadratical coupling strength that can reach $>30\text{MHz}/\text{nm}^2$~\cite{J.C.Sankey@np,D.Lee@nc}; in the superconducting electrical circuit quadratic optomechanical system, the coupling strength $g$ compared to dissipation rate $\kappa$ could be raised to $g/\kappa>0.1$~\cite{E.J.Kim@arxiv}; cavity optomechanical system with ultracold atomic ensembles also provides an obvious characterization of quadratic-coupling regime~\cite{T.P.Purdy@prl}. These platforms offer the possibility to enter the single-photon strong quadratically optomechanical coupling regime.

Here, we investigate the nonlinear mechanical cooling processes in the intrinsic quadratically optomechanical coupling regime without linearizing the optomechanical interaction. We apply the scattering theory to calculate the transition rates between different mechanical Fock states with use of the method of resolvent of Hamiltonian~\cite{M.Bienert@njp,Z.Yi@oc}. This approach allows for a clear identification of the underlying physical processes, where only even-phonon transitions are permitted while the odd-phonon transitions are prohibited. In the paper we first derive the mechanical transition rates, verify the consistence of transition rates with the results of the simulation of full quantum master equation under the consideration of thermal mechanical damping, and then discuss the analytical results in the weak coupling limit that coincide with two-phonon mechanical cooling processes~\cite{A.Nunnenkamp@pra,J.Zhu@jap}. To indicate the difference to the linear coupling optomechanical system by evaluating the Mandel $Q$ parameter, it tends to achieve the nonclassical mechanical states in quadratical optomechanics. Furthermore, we apply the phonon number fluctuations $F$ to present that the even-phonon transitions favor to suppress the phonon number fluctuations.

The paper is organized as follows. In Sec.~\ref{sec2} the quadratical optomechanical system is introduced, in Sec.~\ref{sec3} the scattering theory on the single-photon quadratically optoemchanical cooling process is presented and the cooling limits in the weak and strong coupling regimes are discussed, and in Sec.~\ref{sec4} the nonthermal and nonclassical statistical properties of the mechanical oscillator is analyzed. At last the conclusions are given.

\section{Description of physical model}\label{sec2}
We consider the quadratically coupled optomechanical system in which an optical cavity mode is parametrically coupled to the square of the position of a mechanical oscillator, and the Hamiltonian is written as
\begin{align}
\hat{H}_0=\omega_R\hat{a}^\dag\hat{a}+\omega_m\hat{b}^\dag\hat{b}+g\hat{a}^\dag\hat{a}(\hat{b}^\dag+\hat{b})^2,
\label{eq:H0}
\end{align}
where $\omega_R$ is the frequency of the cavity mode $\hat{a}$, $\omega_m$ is the frequency of the mechanical mode $\hat{b}$, and $g$ is the quadratically optomechanical coupling strength between the cavity mode and mechanical motion of the oscillator. In order to make the present system stable, the Routh-Hurwitz criterion implies that the condition $\omega_m+4sg>0$ should be satisfied, in which $s$ is the number of photons inside the cavity. In the frame rotating at the driving laser's frequency $\omega_L$, the drive is described by the Hamiltonian $\hat{H}^\prime=\Omega(\hat{a}^\dag+\hat{a})$ with the driving strength $\Omega$. We are looking forward to the regime $\Omega\ll\kappa$, where $\kappa$ is the damping rate of the cavity field, and thus the cavity states with more than one photon can be neglected.

The damping of the cavity field is described by the interaction between the mode $\hat{a}$ and the environmental mode $\hat{c}_{k\epsilon}$, where $\hat{H}_\text{emf}=\sum_{k\epsilon}(\omega_k-\omega_L)\hat{c}^\dag_{k\epsilon}\hat{c}_{k\epsilon}$ is the free dynamics of the electromagnetic field external to resonator in the reference frame rotating at the laser frequency $\omega_L$. Here, the sum runs over all modes, identified by the wave vector $k$ and orthogonal polarization $\epsilon$, the operators $\hat{c}_{k\epsilon}$ and $\hat{c}^\dag_{k\epsilon}$ annihilate and create photons in the corresponding mode, and $\omega_k=ck$ denotes the mode frequency. The cavity-environment interaction
\begin{align}
\hat{W}_\kappa=\sum_{k\epsilon}g^{(\kappa)}_{k\epsilon}(\hat{a}^\dag\hat{c}_{k\epsilon}+\hat{a}\hat{c}^\dag_{k\epsilon})
\end{align}
accounts for the coupling of the cavity mode with the modes of the external radiation field and the explicit expressions of coupling constant $g^{(\kappa)}_{k\epsilon}$ can be referred in literatures~\cite{S.M.Dutra,H.J.Carmichael}. The interaction gives the losses of the cavity at rate $\kappa$, with the definition
\begin{align}
\kappa=2\pi\sum_{k\epsilon}\abs{g^{(\kappa)}_{k\epsilon}}^2\delta(\omega_k-\omega_R).
\label{kappa}
\end{align}
The mechanical dissipation rate is usually much smaller than the cavity damping rate, and we will take it into consideration in the following.

\section{Resolvent method on nonlinear cooling processes}\label{sec3}
In this paper we will resort to the scattering theory to explicitly evaluate the transition rates which quantitatively determine the cooling and heating processes. This approach will allow for a direct identification of the underlying physical processes. For convenience we separate the total system Hamiltonian according to $\hat{H}=\hat{H}_0+\hat{V}$, where the redefined main operator $\hat{H}_0$ combines Eq.~\eqref{eq:H0} and free electromagnetic field, i.e.,
\begin{align}
\hat{H}_0=-\Delta\hat{a}^\dag\hat{a}+\omega_m\hat{b}^\dag\hat{b}+g\hat{a}^\dag\hat{a}(\hat{b}^\dag+\hat{b})^2+\hat{H}_\text{emf},
\end{align}
and the small interaction part is
\begin{align}
\hat{V}=\hat{W}_{\kappa}+\hat{H}^\prime,
\label{interaction}
\end{align}
in which $\Delta=\omega_L-\omega_R$ is the detuning between laser $\omega_L$ and resonator frequency $\omega_R$.
In the case of weak optical drive strength $\Omega$, it is sufficient to work to second order of $\Omega$ in the transition rate to well describe the scattering processes. Here we suppose that the initial state is $\ket{i}=\ket{0;n;0_{k\epsilon}}$, where the cavity mode is in the vacuum state $\ket{0}$, $\ket{n}$ is phononic excitations and $\ket{0_{k\epsilon}}$ is environmental vacuum mode. We are interested in the processes that change the motional state of the membrane induced by one photon followed by dissipation to environment, and denote the final states $\ket{f}=\ket{0;m;1_{k\epsilon}}$. Thus we only need to retain the transition amplitude proportional to $\Omega$, which is given by
\begin{align}
\mathcal{T}_\text{fi}=\bra{f}\hat{V}\hat{G}_0(E_i)\hat{V}\ket{i},
\label{TransAmp}
\end{align}
where $\hat{G}_0(z)=1/(z-\hat{H}^\text{eff}_0)$ is the resolvent of the effective Hamiltonian $\hat{H}^\text{eff}_0=\hat{H}_0-i\frac{\kappa}{2}\hat{a}^\dag\hat{a}$ under the consideration of cavity damping, and $E_i=n\omega_m$ is the energy of initial state~\cite{C.Cohen-Tannoudji}. With the help of polaron transformation
\begin{align}
\hat{U}=\exp\left[-\frac{1}{4}\ln\left(1+\frac{4\hat{a}^\dag\hat{a}g}{\omega_m}\right)(\hat{b}^2-\hat{b}^{\dag2})/2\right]
\end{align}
and the relation
\begin{align}
[\hat{U}\frac{1}{z-\hat{H}^\text{eff}_0}\hat{U}^\dag][\hat{U}(z-\hat{H}^\text{eff}_0)\hat{U}^\dag]=\textbf{I},
\end{align}
where $\textbf{I}$ is the identity matrix, the resolvent of the effective Hamiltonian can be reexpressed as
\begin{align}
\hat{G}_0(z)&=\frac{1}{z-\hat{H}^\text{eff}_0}=\hat{U}^\dag\hat{U}\frac{1}{z-\hat{H}^\text{eff}_0}\hat{U}^\dag\hat{U}
\nonumber\\&=\hat{U}^\dag\frac{1}{\hat{U}(z-\hat{H}^\text{eff}_0)\hat{U}^\dag}\hat{U}.
\label{resH0}
\end{align}
By applying the form of polaron transformation, the transformation of the effective Hamiltonian becomes
\begin{align}
\hat{U}(z-\hat{H}^\text{eff}_0)\hat{U}^\dag&=z-\Big[-\Delta\hat{a}^\dag\hat{a}
+\omega_m\sqrt{1+4\hat{a}^\dag\hat{a}g/\omega_m}\nonumber\\&\times(\hat{b}^\dag\hat{b}+1/2)-\frac{1}{2}\omega_m-i\frac{\kappa}{2}\hat{a}^\dag\hat{a}\Big].
\label{unitraytrans}
\end{align}
With the explicit form of interaction part $\hat{V}$ in Eq.~\eqref{interaction} and using the initial state $\ket{i}$ and final state $\ket{f}$, we can find out that the nonzero part of transition amplitude is
\begin{align}
&\mathcal{T}_\text{fi}=\bra{f}\hat{W}_\kappa\hat{G}_0(E_i)\hat{H}^\prime\ket{i}
=\Omega g^{(\kappa)}_{k\epsilon}\sum_{l}\frac{S_{l,m}S_{l,n}}{\delta^{(1)}(l,n)+\Delta+i\frac{\kappa}{2}},\nonumber\\
&\delta^{(1)}(l,n)=(n+\frac{1}{2})\omega_m-(l+\frac{1}{2})\omega^{(1)}_m,\nonumber\\&\omega^{(1)}_m=\omega_m{\sqrt{1+4g/\omega_m}},
\end{align}
in which $S_{l,n}=\bra{l}\exp\left[-\xi(\hat{b}^2-\hat{b}^{\dag2})/2\right]\ket{n}$ with $\xi=\frac{1}{4}\ln(1+4g/\omega_m)$ is the coefficient of squeezed number state in Fock representation and defined as~\cite{M.S.Kim@pra,J.Q.Liao@pra,J.Q.Liao@SR}
\begin{align}
S_{l,n}&=\frac{\sqrt{l!n!}}{(\cosh\xi)^{n+1/2}}\sum_{k^\prime=0}^{\text{Floor}[\frac{l}{2}]}\sum_{k=0}^{\text{Floor}[\frac{n}{2}]}
\frac{(-1)^k}{k^\prime!k!}\nonumber\\&\times\frac{(\frac{1}{2}\tanh\xi)^{k^\prime+k}}{(n-2k)!}(\cosh\xi)^{2k}\delta_{l-2k^\prime,n-2k},
\end{align}
where the function Floor[$x$] gives the greatest integer less than or equal to $x$, and $\omega^{(1)}_m$ is the one-photon coupled membrane's resonant frequency~\cite{H.Shi@pra}.

Then the corresponding transition rate is defined as
\begin{align}
R_\text{fi}=2\pi\sum_{k\epsilon}\abs{\mathcal{T}_\text{fi}}^2\delta(E_i-E_f),
\end{align}
where $\delta$-function guarantees the energy conservation and the sum covers all relevant polarizations and wave vectors. The transition rate becomes
\begin{align}
R_\text{fi}=&2\pi\Omega^2\sum_{k\epsilon}\abs{g^{(\kappa)}_{k\epsilon}}^2\nonumber\\&\times\delta(\omega_k-\omega_R)\abs{\sum_{l}\frac{S_{l,m}S_{l,n}}{\delta^{(1)}(l,n)+\Delta+i\frac{\kappa}{2}}}^2.
\end{align}
With the definition of cavity damping in Eq.~\eqref{kappa}, the transition rate which denotes the change of the motional state of mechanical oscillator from $n$ to $m$ phonons becomes
\begin{align}
R_{fi}=\Gamma_{n\rightarrow m\neq n}=\kappa\Omega^2\abs{\sum_{l}\frac{S_{l,m}S_{l,n}}{\delta^{(1)}(l,n)+\Delta+i\frac{\kappa}{2}}}^2.
\label{Transrate}
\end{align}
Equation~\eqref{Transrate} presents a clear physical view on the process of an incident photon scattered by the quadratic cavity-optomechanical system with changing the motion of mechanical oscillator from $n$ to $m$ phonons. When a laser photon turns into the cavity photon, the phonon state changes from the initial number $n$ to an intermediate number $l$ demanding that $\abs{l-n}$ is even due to the term of two-phonon operators, and the amplitude is inversely proportional to the frequency detuning $\delta^{(1)}(l,n)+\Delta$ which relates to the photon-coupled membrane's resonant frequency $\omega_m^{(1)}$, and proportional to the matrix element of squeezed number state $S_{l,n}$. As the photon leaves the cavity and dissipates into the continuum of modes in free space, it induces a transition in mechanical motion from $l$ to $m$ phonons which also demands that $\abs{l-m}$ is even, and its amplitude is independent of driving detuning but dependent on the cavity-reservoir coupling strength $g^{(\kappa)}_{k\epsilon}$, and also determined by the coefficient of squeezed number state $S_{l,m}$ for the two-phonon term. The output photon possesses the energy $\omega_L+(n-m)\omega_m$ and carries away $(n-m)$ phonons, in which $(n-m)$ should be even since $\abs{l-n}$ and $\abs{l-m}$ are both even.

Now we stress the differences of nonlinear cooling rates between quadratical and linear coupling optomechanical systems. The quadratical optomechanical coupling modifies the mechanical frequency of the oscillator while the linear coupling just displaces the mechanical equilibrium position. Thus the frequency detuning of the transition $\delta^{(1)}(l,n)$ is closely dependent on the coupling strength $g$ while the effective detuning in the linear coupling optomechanics is independent of the coupling strength in the single-photon level. Moreover, the transition amplitude of the linear coupling optomechanics is proportional to the Franck-Condon overlap factor for the single-phonon operators while the transition amplitude is proportional to the matrix elements of squeezed number state for the two-phonon operators. Therefore, the form of transition rate in Eq.~\eqref{Transrate} is different from that of linear coupling optomechanical system in Ref.~\cite{Z.Yi@oc,A.Nunnenkamp@praR}, and may be helpful to generate more interesting mechanical states.

Taking into account of the thermal damping of mechanical oscillator with rate $\gamma_m$ and thermal phonon number $n_\text{th}$, the set of rate equations for the mechanical oscillator becomes
\begin{align}
\dot{P}_n=&-\gamma_m n_\text{th}(n+1)P_n-\gamma_m(n_\text{th}+1)nP_n\nonumber\\&+\gamma_m n_\text{th}nP_{n-1}+\gamma_m(n_\text{th}+1)(n+1)P_{n+1}\nonumber\\&-\sum_{m\neq n}\Gamma_{n\rightarrow m}P_n+\sum_{m\neq n}\Gamma_{m\rightarrow n}P_m, (\abs{m-n} \text{is even})
\label{eq:rate}
\end{align}
where $P_n$ is the phonon number distribution in the $n$ phonon state.
\subsection{Cooling process in the resolved-sideband and weak quadratical coupling limits}
The occurrence of cooling processes in the strong quadratical coupling regime can absorb multiple two-phonons, and modify the eigenfrequency of mechanical oscillator simultaneously, which can lead to the generation of uncommon mechanical states. To understand this behavior we first consider the case of resolved-sideband limit $\omega_m\gg\kappa$ and weak quadratical coupling limit $g\ll\omega_m$, where the scattering is mainly dominated by the two-phonon transitions $\Gamma_{n\rightarrow n\pm2}$, and the rates in Eq.~\eqref{Transrate} are explicitly simplified as $\Gamma_{n\rightarrow n-2}=n(n-1)\Gamma_\downarrow$ and $\Gamma_{n\rightarrow n+2}=(n+1)(n+2)\Gamma_\uparrow$ in these limits, in which
\begin{align}
\Gamma_\downarrow&=\frac{\kappa\Omega^2g^2/\omega_m^2}{(\kappa/2)^2+(\Delta+2\omega_m)^2},\nonumber\\
\Gamma_\uparrow&=\frac{\kappa\Omega^2g^2/\omega_m^2}{(\kappa/2)^2+(\Delta-2\omega_m)^2}
\end{align}
describe the strengths of two-phonon absorption and emission processes. The results are consistent with effective two-phonon quadratically mechanical cooling theory~\cite{A.Nunnenkamp@pra}, in which the cavity photon is linearized and then served as the two-phonon reservoir in the weak optomechanical coupling limit. The set of rate equation in Eq.~\eqref{eq:rate} becomes
\begin{align}
\dot{P}_n=&\gamma_m(n_\text{th}+1)[(n+1)P_{n+1}-nP_n]\nonumber\\&-\gamma_m n_\text{th}[(n+1)P_n-nP_{n-1}]\nonumber\\&
+\Gamma_\downarrow[(n+1)(n+2)P_{n+2}-n(n-1)P_n]\nonumber\\&-\Gamma_\uparrow[(n+1)(n+2)P_n-n(n-1)P_{n-2}].
\label{eq:srate}
\end{align}

For the high-Q mechanical oscillator and low initial temperature, i.e., $\gamma_mn_\text{th}\ll\Gamma_{\downarrow\uparrow}$, we can ignore the thermal damping effects. The cooling process turns to be the pure two-phonon transitions, and the phonon number distribution $P_n$ can be analytically solved and expressed in the form
\begin{align}
P^{(2)}_{2n+j}=(1-r)r^{n}(\gamma+j-1)(-1)^{j-1},\hspace{5pt}j=0,1,
\end{align}
where $r=\Gamma_\uparrow/\Gamma_\downarrow$, and $\gamma$ characterizes the relative weight of the odd phonon states determined by the initial conditions~\cite{V.V.Dodonov@jpa}. In the resolved-sideband regime, when the detuning satisfies $\Delta=-2\omega_m$, the two-phonon absorption process dominates the emission process, i.e., $r\ll1$, and the phonon number distribution is concentrated at zero- and one-phonon states $P_0=1-\gamma$, $P_1=\gamma$, which indicates that two-phonon cooling processes preserve the phonon-number parity, leading to the initial odd phonon states cooled to the one-phonon state and the even phonon states cooled to the zero-phonon state independently.

In the strong two-phonon absorption regime $\Gamma_\downarrow\gg\gamma n_\text{th}\gg\Gamma_\uparrow$, which is achievable with $\Delta=-2\omega_m$ in the resolved-sideband limit, finally we can obtain~\cite{V.V.Dodonov@jpa}
\begin{align}
P_0=&\frac{1+2\xi}{1+3\xi}+\mathcal{O}(\frac{\gamma_mn_\text{th}}{\Gamma_\downarrow}),\nonumber\\
P_1=&\frac{\xi}{1+3\xi}+\mathcal{O}(\frac{\gamma_mn_\text{th}}{\Gamma_\downarrow}),\hspace{5pt} \xi=\frac{n_\text{th}}{n_\text{th}+1}.
\label{eq:weakpopul}
\end{align}
Thus the minimal mean phonon number $\langle\hat{n}\rangle=1/(4+1/n_\text{th})$, and the Mandel $Q$ factor that is defined as $Q=\frac{\langle\hat{n}^2\rangle-\langle\hat{n}\rangle^2}{\langle\hat{n}\rangle}-1$ with $\langle\hat{n}\rangle=\sum_{n=0}^\infty nP_n, \langle\hat{n}^2\rangle=\sum_{n=0}^\infty n^2P_n$ becomes $Q\approx-\langle\hat{n}\rangle$, which shows sub-Possionian phonon distribution. Thus the membrane is in a nonclassical mechanical state.

\subsection{Cooling process in the strong coupling limit}
\begin{figure}[hbt]
\centering\includegraphics[width=8cm,keepaspectratio,clip]{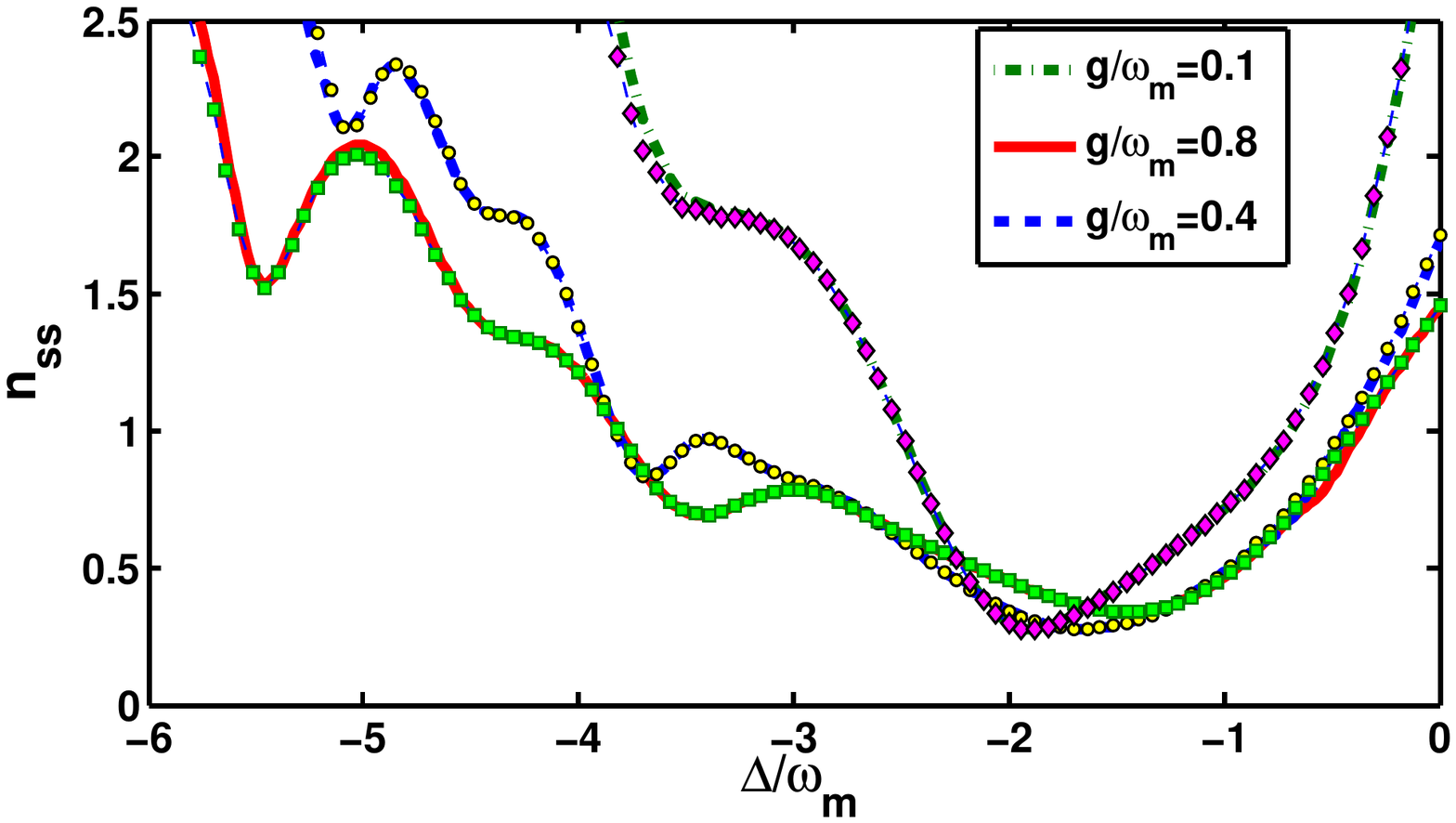}
\caption{ (Color online) Steady-state phonon number $n_{ss}=\langle\hat{b}^\dag\hat{b}\rangle$ in single-photon quadratical optomechanical system as a function of detuning $\Delta$ with the coupling strengths $g/\omega_m=0.8$ (red solid line and green squares), $g/\omega_m=0.4$ (blue dashed line and yellow circles) and $g/\omega_m=0.1$ (green dash-dotted line and magenta diamonds). The other parameters are $\gamma_m/\omega_m=10^{-6}$, $n_\text{th}=10$, $\kappa/\omega_m=0.25$ and $\Omega/\kappa=0.4$. (In the figure the lines are obtained from the simulation of the master equation~\eqref{eq:master} by using the Quantum Optics Toolbox and the markers are obtained by solving the rate equations in Eq.~\eqref{eq:rate}.)}
\label{fig:nphonon}
\end{figure}

In the strong coupling regime higher-order mechanical sidebands appear and the multiple two-phonon transitions begin to work. In Fig.~\ref{fig:nphonon} we plot the steady-state phonon number $n_{ss}=\langle\hat{b}^\dag\hat{b}\rangle$ as a function of detuning $\Delta=\omega_L-\omega_R$ with different coupling strengths $g/\omega_m$. We observe that in the weak coupling and the resolved-sideband limits, i.e., $g/\omega_m=0.1$, the optimal cooling occurs at the detuning $\Delta=-2\omega_m$ and the cooling limit reaches the value close to $0.25$, which is in accordance to Eq.~\eqref{eq:weakpopul}. In the strong coupling limit, the efficient cooling processes occur at the multiple two-phonon resonances $\Delta=-\delta^{(1)}(l,n)$, where the multiple dips of the evolution of $n_{ss}$ indicate the occurrence of the multiple two-phonon transitions, which will further influence the statistical properties of the mechanical oscillator.

To indicate the validity of the set of rate equations obtained by scattering theory in Eq.~\eqref{eq:rate}, we compare the steady-state phonon number with numerically solutions of the quantum master equation
\begin{align}
\frac{d}{dt}\hat{\rho}&=-i[\tilde{\hat{H}},\hat{\rho}]+\frac{\kappa}{2}\mathcal{L}[\hat{a}]\hat{\rho}+\gamma_m(n_\text{th}+1)\mathcal{L}[\hat{b}]\hat{\rho}\nonumber\\&+\gamma_mn_\text{th}\mathcal{L}
[\hat{b}^\dag]\hat{\rho},
\label{eq:master}
\end{align}
performed with the Quantum Optics Toolbox~\cite{S.M.Tan}, where the Hamiltonian is $\tilde{\hat{H}}=-\Delta\hat{a}^\dag\hat{a}+\omega_m\hat{b}^\dag\hat{b}+g\hat{a}^\dag\hat{a}(\hat{b}+\hat{b}^\dag)^2+\Omega(\hat{a}+\hat{a}^\dag)$, $\mathcal{L}[\hat{o}]\hat{\rho}=\hat{o}\hat{\rho}\hat{o}^\dag-\frac{1}{2}\{\hat{o}^\dag\hat{o},\hat{\rho}\}$ is the Lindblad operator of photon (phonon) dissipation. In Fig.~\ref{fig:nphonon}, we plot the results of master equation~\eqref{eq:master} in variant style of lines, which are respectively red solid, blue dashed and green dash-dotted lines, for different coupling strengths $g/\omega_m=0.8, 0.4, 0.1$, and specific markers, which are green squares, yellow circles and magenta diamonds, are obtained by solving the set of rate equations~\eqref{eq:rate}. These two results are well matched, the rate equation approach coincides with the master equation, and thus scattering theory is adoptable in describing nonlinear quadratical cooling processes.

\section{Statistical properties of steady mechanical state}\label{sec4}
The occurrence of the multiple two-phonon transitions in cooling processes, especially in the strong coupling regime, decreases the membrane's vibrating energy containing several two-phonons simultaneously, and together with the modification on resonant frequency of membrane due to the quadratical coupling, the mechanical oscillator can present different statistical properties compared with linear coupling and linearized optomechanics. In the section we mainly resort to Mandel $Q$ parameter and phonon number fluctuations $F$ to discuss the mechanical statistical properties.

\begin{figure}[hbt]
\centering\includegraphics[width=8cm,keepaspectratio,clip]{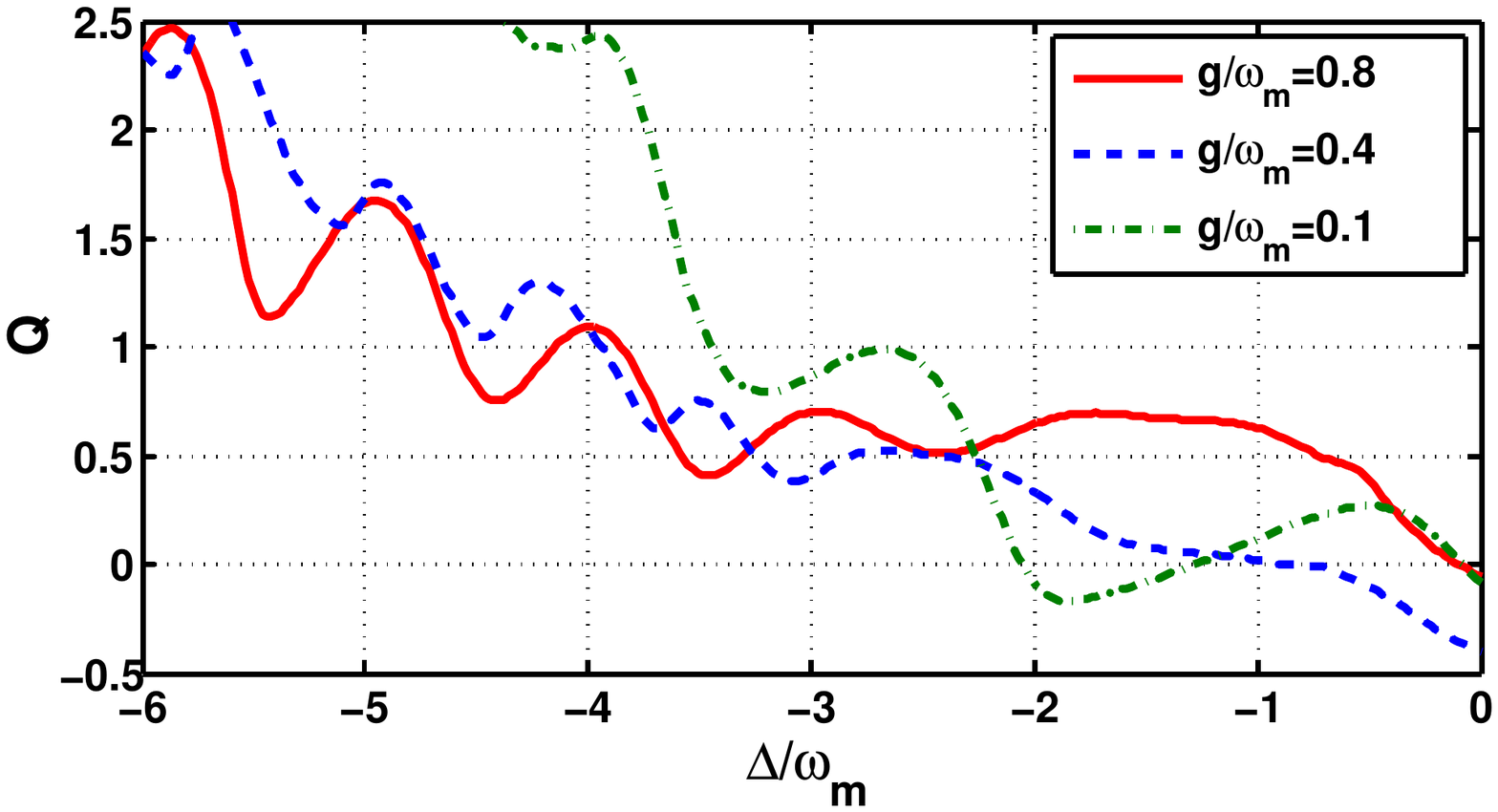}
\caption{ (Color online) Mandel $Q$ parameter in single-photon quadratical optomechanical system as a function of detuning $\Delta$ with the coupling strengths $g/\omega_m=0.8$ (red solid line), $g/\omega_m=0.4$ (blue dashed line) and $g/\omega_m=0.1$ (green dash-dotted line). The other parameters are $\gamma_m/\omega_m=10^{-6}$, $n_\text{th}=10$, $\kappa/\omega_m=0.25$ and $\Omega/\kappa=0.4$.}
\label{fig:Q}
\end{figure}

By using Mandel $Q$ parameter it is convenient to characterize nonclassical states with negative values, which indicate a sub-Poissonian statistics and have non classical analog. In Fig.~\ref{fig:Q}, we plot the Mandel $Q$ parameter as a function of detuning with different quadratical coupling strengths. In the weak coupling regime, i.e. $g/\omega_m=0.1$, the minimum value of $Q$ becomes negative that is close to the value $-\langle\hat{n}\rangle$, which agrees with the results of Eq.~\eqref{eq:weakpopul}. With the increasing coupling strength, $Q$ can present the negative value as well. For example, at $g/\omega_m=0.4$ it takes a smaller value. Thus the mechanical oscillator is in a nonclassical state. However, in the strong linear coupling optomechanics~\cite{A.Nunnenkamp@praR,Z.Yi@oc}, it is relatively difficult to present the nonclassical properties in weak driving regime.

\begin{figure}[hbt]
\centering\includegraphics[width=8cm,keepaspectratio,clip]{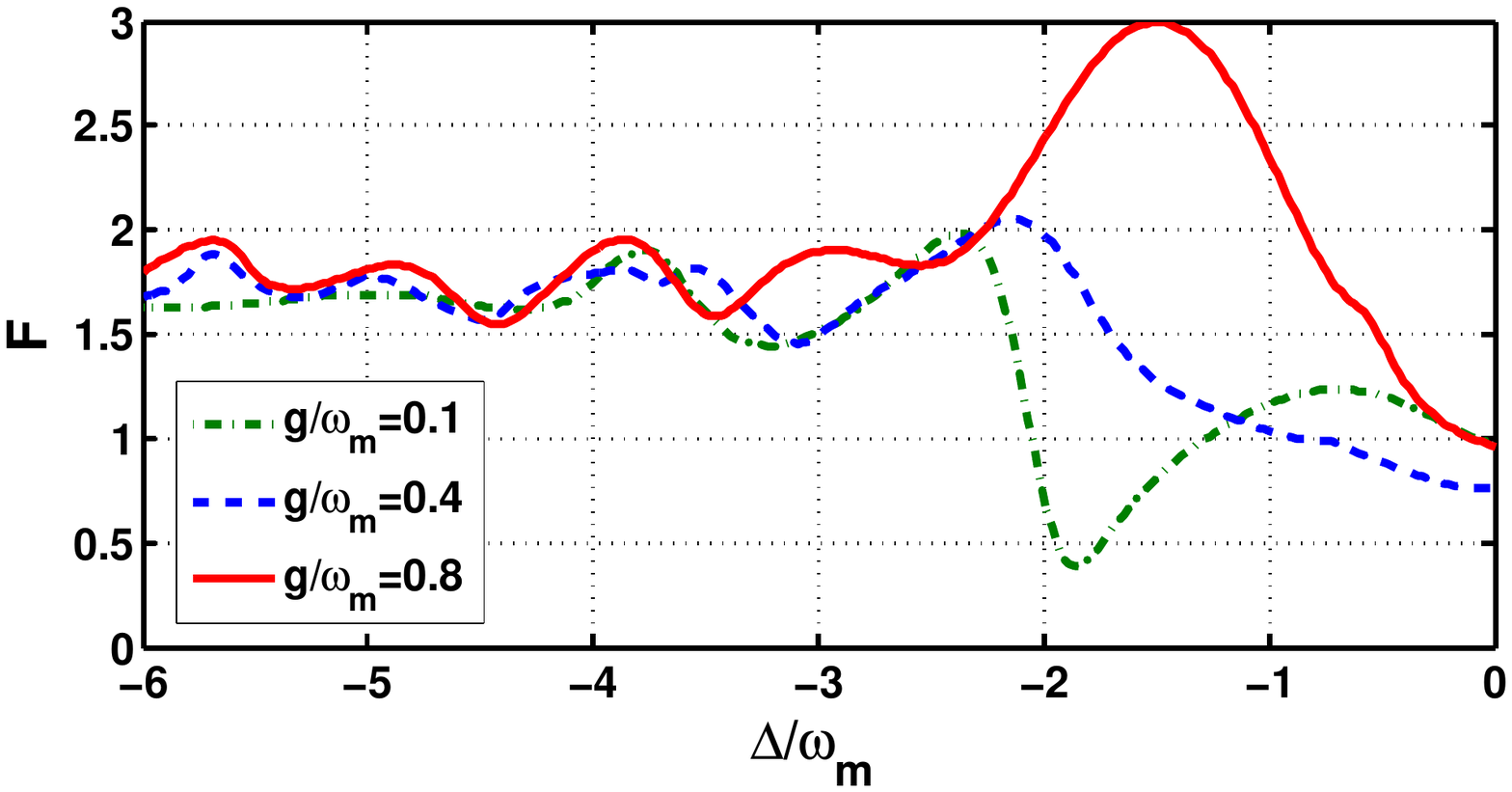}
\caption{ (Color online) Phonon number fluctuations F in single-photon quadratical optomechanical system as functions of detuning $\Delta$ with the coupling strengths $g/\omega_m=0.8$ (red solid line), $g/\omega_m=0.4$ (blue dashed line) and $g/\omega_m=0.1$ (green dash-dotted line). The other parameters are $\gamma_m/\omega_m=10^{-6}$, $n_\text{th}=10$, $\kappa/\omega_m=0.25$ and $\Omega/\kappa=0.4$.}
\label{fig:F}
\end{figure}

In the linearized treatment of mechanical cooling regime, the rate equation satisfies the detailed balance condition which implies that the final phonon state is in a thermal state. For the thermal state, $\langle\hat{n}\rangle=n_\text{th}$, $\langle\hat{n}^2\rangle=2n_\text{th}^2+n_\text{th}$, yielding the phonon number fluctuations $F=\langle\hat{b}^\dag\hat{b}^\dag\hat{b}\hat{b}\rangle/\langle\hat{b}^\dag\hat{b}\rangle^2=2$~\cite{S.Lee}. Here the rate equations~\eqref{eq:rate} obviously do not obey the detailed balance condition and the steady mechanical state will be in a nonthermal state. It is more clearly exhibited in the single two-phonon cooling process of the weak coupling limit in Eq.~\eqref{eq:srate}. In Fig.~\ref{fig:F} we plot the phonon number fluctuations $F$ as functions of detuning $\Delta$ with different coupling strengths. Since phonon number fluctuations $F=2$ for the thermal mechanical state, in the nonlinear intrinsic quadratical optomechanics we can both decrease and increase the number fluctuations ($F<2$ and $F>2$), and the fluctuations can be suppressed for a large range of detuning. Moreover, compared with the optomechanical cooling induced by single-photon linear coupling where the fluctuation is mainly enlarged while the mechanical cooling occurs here with the mainly decreased phonon number fluctuations, which possesses potential applications. The suppression is induced by the (multiple) two-phonon cooling processes that decrease the phonon occupation in the zero- and one-phonon number states, and compared to one-phonon cooling processes it is more achievable to suppress the phonon number fluctuations~\cite{A.Nunnenkamp@praR,A.Nunnenkamp@pra}.

\section{Conclusions}\label{sec5}
In conclusion, we have studied the nonlinear mechanical cooling processes in the intrinsic quadratically optomechanical coupling regime without linearizing the optomechanical interaction. We apply the scattering theory to calculate the transition rates between different mechanical Fock states with the use of the method of resolvent of Hamiltonian, since the approach can present a direct identification of the underlying physical processes. Due to the preservation of phonon number parity, only even-phonon transitions are permitted and the odd-phonon transitions are forbidden. In the paper we first derive the phononic transition rates, and verify the feasibility of scattering approach by comparing with the results of simulation of the full quantum master equation. We also discuss the analytical mechanical cooling limits in the weak coupling limit, and find that they coincide with two-phonon mechanical cooling processes. Finally, the statistical properties of the mechanical state is presented. The mechanical state can be in a sub-Poissonian distribution that characterizes the nonclassical state, and the even-phonon transitions favor to suppress the phonon number fluctuations for a large range of detuning as well.

{\bf{Acknowledgements:}}
The authors gratefully acknowledge the support by Startup Foundation for Doctors of Yangtze University (Grant No. 8010800101), the National Natural Science Foundation of China (Grant No. 11304024) and the Technology Creative Project of Excellent Middle and Young Team of Hubei Province (Grant No. T201204).


\begin{thebibliography}{99}
\bibitem{Sh.barzanjeh@prl} Sh. Barzanjeh, M. Abdi, G. J. Milburn, P. Tombesi, and D. Vitali, Phys. Rev. Lett. {\bf 109}, 130503 (2012).
\bibitem{V.Fiore@pra} V. Fiore, Y. Yang, M. C. Kuzyk, R. Barbour, L. Tian, and H. Wang, Phys. Rev. Lett. {\bf 107}, 133601 (2011).
\bibitem{C.Dong@sci} C. Dong, V. Fiore, M. C. Kuzyk, H. Wang, Science {\bf 338}, 1609-1613 (2012).
\bibitem{J.Li@pr} J. J. Li and K. D. Li, Phys. Rep. {\bf 525}, 223-254 (2013).
\bibitem{T.J.Kippenberg@sci} T. J. Kippenberg and K. J. Vahala, Science {\bf 321}, 1172-1176 (2008).
\bibitem{A.A.Gangat@pra} A. A. Gangat, Phys. Rev. A {\bf 88}, 063846 (2013).
\bibitem{I.Wilson-Rae@prl} I. Wilson-Rae, N. Nooshi, W. Zwerger, and T. J. Kippenberg, Phys. Rev. Lett. {\bf 99}, 093901 (2007).
\bibitem{J.D.Teufel@nat} J. D. Teufel, T. Donner, D. Li, J. W. Harlow, M. S. Allman, K. Cicak, A. J. Sirois, J. D. Whittaker, K. W. Lehnert, and R. W. Simmonds, Nature {\bf 475}, 359-363 (2011).
\bibitem{W.Gu@pra} W. Gu and G. Li, Phys. Rev. A {\bf 87}, 025804 (2013).
\bibitem{R.L.de Matos Filho} R. L. de Matos Filho and W. Vogel, Phys. Rev. A {\bf 50}, R1988 (1994).
\bibitem{A.Nunnenkamp@praR} A. Nunnenkamp, K. B{\o}rkje, and S. M. Girvin, Phys. Rev. A {\bf 85}, 051803(R) (2012).
\bibitem{M.Koch@prl} M. Koch, C. Sames, M. Balbach, H. Chibani, A. Kubanek, K. Murr, T. Wilk, and G. Rempe, Phys. Rev. Lett. {\bf 107}, 023601 (2011).
\bibitem{X.Lu@prl} X. L\"{u}, Y. Wu, J. R. Johansson, H. Jing, J. Zhang, and F. Nori, Phys. Rev. Lett. {\bf 114}, 093602 (2015).
\bibitem{J.D.Thompson@nat} J. D. Thompson, B. M. Zwickl, A. M. Jayich, F. Marquardt, S. M. Girvin, and J. G. E. Harris, Nature {\bf 452}, 72-75 (2008).
\bibitem{T.P.Purdy@prl} T. P. Purdy, D. W. C. Brooks, T. Botter, N. Brahms, Z. Y. Ma, and D. M. Stamper-Kurn, Phys. Rev. Lett. {\bf 105}, 133602 (2010).
\bibitem{E.J.Kim@arxiv} E. J. Kim, J. R. Johansson, and F. Nori, arXiv:1412.6869.
\bibitem{J.Q.Liao@pra} J.-Q. Liao and F. Nori, Phys. Rev. A {\bf 88}, 023853 (2013).
\bibitem{J.Q.Liao@SR} J.-Q. Liao and F. Nori, Sci. Rep. {\bf 4}, 06302 (2014).
\bibitem{H.Shi@pra} H. Shi and M. Bhattacharya, Phys. Rev. A {\bf 87}, 043829 (2013).
\bibitem{A.A.Clerk@prl} A. A. Clerk, F. Marquardt, and J. G. E. Harris, Phys. Rev. Lett. {\bf 104}, 213603 (2010).
\bibitem{M.Asjad@pra} M. Asjad, G. S. Agarwal, M. S. Kim, P. Tombesi, G. D. Giuseppe, and D. Vitali, Phys. Rev. A {\bf 89}, 023849 (2014).
\bibitem{S.Huang@pra} S. Huang and G. S. Agarwal, Phys. Rev. A {\bf 83}, 023823 (2011).
\bibitem{Z.J.Deng@pra} Z. J. Deng, Y. Li, M. Gao, and C. W. Wu, Phys. Rev. A {\bf 85}, 025804 (2012).
\bibitem{J.C.Sankey@np} J. C. Sankey, C. Yang, B. M. Zwickl, A. M. Jayich, and J. G. E. Harris, Nature Phys. {\bf 6}, 707-712 (2010).
\bibitem{D.Lee@nc} D. Lee, M. Underwood, D. Mason, A. B. Shkarin, S. W. Hoch, and J. G. E. harris, Nat. Commun. {\bf 6}, 6232 (2015).
\bibitem{M.Bienert@njp} M. Bienert and G. Morigi, New J. Phys. {\bf 14}, 023002 (2012).
\bibitem{Z.Yi@oc} Z. Yi, W. Gu, S. Wei, and D. Xu, Opt. Commun. {\bf 341}, 28-31 (2015).
\bibitem{A.Nunnenkamp@pra} A. Nunnenkamp, K. B{\o}rkje, J. G. E. Harris, and S. M. Girvin, Phys. Rev. A {\bf 82}, 021806(R) (2010).
\bibitem{J.Zhu@jap} J.Zhu and G. Li, J. Appl. Phys. {\bf 111}, 033704 (2012).
\bibitem{S.M.Dutra} S. M. Dutra and G. Nienhuis, J. Opt. B {\bf 2}, 584 (2000).
\bibitem{H.J.Carmichael} H. J. Carmichael, \emph{Statistical Methods in Quantum Optics 1} (Springer, New York, 2002).
\bibitem{C.Cohen-Tannoudji} C. Cohen-Tannoudji, J. Dupont-Roc and G. Grynberg, \emph{Atom-Photon Interactions (Basic Processes and Applications)}.
\bibitem{M.S.Kim@pra} M. S. Kim, F. A. M. de Oliveira, and P. L. Knight, Phys. Rev. A {\bf 40}, 2494 (1989).
\bibitem{V.V.Dodonov@jpa} V. V. Dodonov and S. S. Mizrahi, J. Phys. A: Math. Gen. {\bf 30} 5657-5667 (1997).
\bibitem{S.M.Tan} S. M. Tan, J. Opt. B {\bf 1}, 424 (1999).
\bibitem{S.Lee} S. Lee, J. Park, S. Ji, C. H. R. Ooi, and H. Lee, J. Opt. Soc. Am. B {\bf 26}, 1532-1537 (2009).





\end{thebibliography}
\end{document}